\documentclass[conference]{IEEEtran}
\IEEEoverridecommandlockouts
\usepackage{amsmath,amssymb,amsfonts}
\usepackage{algorithmic}
\usepackage{graphicx}
\usepackage{textcomp}
\usepackage{xcolor}

\usepackage{lipsum}
\usepackage{multirow}
\usepackage{subcaption}
\usepackage{tikz}
\usepackage{balance}
\usepackage{enumitem}
\usepackage{tablefootnote}
\usepackage{algorithm}
\usepackage{listings}
\usepackage{makecell}
\usepackage{marvosym}

\def\BibTeX{{\rm B\kern-.05em{\sc i\kern-.025em b}\kern-.08em
    T\kern-.1667em\lower.7ex\hbox{E}\kern-.125emX}}
\begin{document}

\title{CoroAMU: Unleashing Memory-Driven Coroutines through Latency-Aware Decoupled Operations\\
\thanks{\Letter\ Corresponding Author}
}

\author{
\IEEEauthorblockN{
Zhuolun Jiang\IEEEauthorrefmark{1}\IEEEauthorrefmark{2},
Songyue Wang\IEEEauthorrefmark{1}\IEEEauthorrefmark{2},
Xiaokun Pei\IEEEauthorrefmark{1}\IEEEauthorrefmark{2},
Tianyue Lu\IEEEauthorrefmark{1}\IEEEauthorrefmark{2}\Letter\ and
Mingyu Chen\IEEEauthorrefmark{1}\IEEEauthorrefmark{2}}

\IEEEauthorblockA{\IEEEauthorrefmark{1}State Key Lab of Processors\\
Institute of Computing Technology, Chinese Academy of Sciences
}
\IEEEauthorblockA{\IEEEauthorrefmark{2}University of Chinese Academy of Sciences
}
\IEEEauthorblockA{\{jiangzhuolun22z, wangsongyue22s, peixiaokun23z, lutianyue, cmy\}@ict.ac.cn}
}

\maketitle

\begin{abstract}
Modern data-intensive applications face memory latency challenges exacerbated by disaggregated memory systems. Recent work shows that coroutines are promising in effectively interleaving tasks and hiding memory latency, but they struggle to balance latency-hiding efficiency with runtime overhead. We present CoroAMU, a hardware-software co-designed system for memory-centric coroutines.
It introduces compiler procedures that optimize coroutine code generation, minimize context, and coalesce requests, paired with a simple interface. With hardware support of decoupled memory operations, we enhance the Asynchronous Memory Unit to further exploit dynamic coroutine schedulers by coroutine-specific memory operations and a novel memory-guided branch prediction mechanism. It is implemented with LLVM and open-source XiangShan RISC-V processor over the FPGA platform. Experiments demonstrate that the CoroAMU compiler achieves a 1.51× speedup over state-of-the-art coroutine methods on Intel server processors. When combined with optimized hardware of decoupled memory access, it delivers 3.39× and 4.87× average performance improvements over the baseline processor on FPGA-emulated disaggregated systems under 200ns and 800ns latency respectively.
\end{abstract}


\section{Introduction} \label{sec:intro}

Modern data-center applications increasingly rely on manipulating massive in-memory datasets, such as graph processing\cite{intro_graph_ligra_13ppopp}\cite{intro_graph_gemini_16osdi}, in-memory databases\cite{intro_database_1_13SIGMOD}\cite{intro_database_2_16SIGMOD}\cite{intro_database_3_22SIGMOD}, deep learning\cite{intro_dl_pytorch_19nips}, and scientific computing. The inherent parallelism in these applications—whether data-level or task-level—has driven the adoption of multi-core systems to exploit memory bandwidth\cite{intro_scaleout_12isca} and improve memory utilization. Yet frequent memory accesses leave individual CPU cores underutilized, trapped in stalls while waiting for data, resulting in excessive energy consumption\cite{intro_memorywall_95}\cite{motivation_memory_99vldb}\cite{intro_memory_cloud_17}.

To further address the challenge of low memory utilization at the data-center level\cite{intro_datacenter_mem_22asplos}, modern systems increasingly adopt disaggregated memory\cite{lim2012hpca}, where applications can leverage remote memory resources over network interconnects (e.g., CXL\cite{CXL}, RDMA\cite{rdma}). Software-defined far memory\cite{intro_farmem_19asplos}\cite{intro_farmem_2_24osdi} and cloud-native systems\cite{PolarDB-MP_2024icde}\cite{intro_cloud_24} are gaining popularity and introducing new critical problems: extended access latency, unpredictable multi-hop network delays, and inconsistent latency across distributed nodes. The end-to-end latency in CXL-enabled disaggregated memory systems reaches over 300ns as reported by \cite{li2023asplos}. 

Researchers have explored software techniques with task interleaving and software prefetching to mitigate memory latency\cite{amac}\cite{clairvoyance}. It extracts parallel instruction streams from programs, issuing \texttt{prefetch} instructions for long-latency memory operations before switching execution to alternate streams. While effective in practice, these approaches impose significant burdens on programmers.
Recent work replaces manual code modifications with coroutines\cite{c++20coroutine}, introduced in C++20 standard, by allowing functions to suspend and resume execution voluntarily, exhibiting great potential\cite{prefetch_sw_coro}\cite{decouple_cimple}\cite{corobase}\cite{corograph}. However, the use of coroutines in memory-intensive applications remains limited to simple data structures or specialized execution engines. Converting large codebases to coroutines requires intrusive modifications, including identifying all potential suspension points and encapsulating relevant functions, prone to human error. Moreover, while coroutines excel at masking long-latency operations like I/O, where user-mode switching overhead is negligible, their performance impact in short-latency memory-driven scenarios has not been fully analyzed.

A deeper challenge lies in scheduling mechanisms for coroutine-based latency hiding. In traditional architecture, the CPU has no active means to know whether prefetched data has been delivered into the cache. Existing approaches predominantly rely on static scheduling. 
It enforces a deterministic task launch sequence with ordered resumption to precisely regulate prefetch-to-use latencies for individual memory objects.
At the same time, it faces scalability issues. CoroGraph\cite{corograph} reported performance degradation when using more than two coroutines, as fixed task sequences struggle to balance resource contention and prefetch accuracy. Static schedules also cannot adapt to variable network or memory latencies.

Dynamic scheduling can address these limitations, enabled by decoupled memory operations to monitor memory event status and schedule corresponding coroutines. It requires hardware support for two fundamental operations: (1) initiating memory access requests and (2) querying their completion status. Prior work has demonstrated viable implementations through prefetch and cache check-miss instructions as in \cite{decouple_swoop}, decoupled memory request issuing (\texttt{aload} \& \texttt{astore}) and response retrieving (\texttt{getfin}) instructions in AMU\cite{wang2022carrv}\cite{amu}, or via on-chip accelerators with launch-and-poll interfaces\cite{rocc}\cite{intro_acc_21micro}\cite{intro_acc_23micro}. By allowing coroutines to resume only when their prefetched data is ready, dynamic scheduling adapts seamlessly to variable latencies. But this flexibility comes at a cost: resolving indirect jumps between coroutines introduces unpredictable branches, and compilers lack optimizations to mitigate coroutine switching costs, constraining its practical deployment in high-performance systems.

Our goal is to investigate an effective approach to leverage coroutines for hiding memory access latency, which should support adaptive handling of both long and variable delays while maintaining low switching overhead. Crucially, we aim to design an interface that requires little programmer intervention. Our work builds upon the following key insight: Although coroutine-style code generation inevitably introduces additional control overheads, 
there is room for optimization if we know in advance that every switch is due to memory access operations.

First, memory-intensive parallel tasks often exhibit single-source homogeneity (e.g., OpenMP for-loops\cite{openmp}), in contrast to coarse-grained tasks such as I/O operations, where CPU workloads typically invoke distinct processing functions and execute heterogeneous control flows based on data content and sources. Current compilers like LLVM\cite{llvm} make no assumptions about this characteristic. Their generated code conventionally switches between schedulers and tasks through function calls and pointer passing. 
Moreover, shared data structures among coroutines are common. Whenever accesses to such data are separated by suspension points, current compilers create duplicate copies.

Second, the frequency of coroutine switching directly correlates with the number of memory accesses in applications. By coalescing memory requests, switching occurrences can be effectively reduced. Our compiler identifies two coalescing opportunities: (1) memory requests exhibiting spatial locality, and (2) independent memory requests without data dependencies. While request coalescing is straightforward for software prefetching, where overly aggressive merging only impacts performance rather than correctness, dynamic scheduling of coalesced requests for decoupling hardware demands careful algorithm design and hardware support.

Third, hardware-assisted dynamic scheduling achieves adaptive latency hiding but introduces unpredictable indirect jumps determined by external responses. We observe that when the CPU's branch predictor attempts to predict a taken branch, the external response must have arrived at the core. These jump targets can be embedded within the memory responses and injected into the branch predictor, guaranteeing consistently accurate predictions.

On the software side, we design and implement an LLVM-based coroutine compiler optimized for memory latency. Furthermore, for hardware, we enhance AMU as an example of architectural support of efficient dynamic scheduling with decoupled operations.

Our key contributions are as follows:

\begin{itemize}[leftmargin=*]

\item We point out that SOTA coroutine does not adequately consider memory-centric characteristics, quantifying both its effectiveness for memory latency hiding and overheads.

\item We present a set of novel coroutine-based compiler optimization passes targeting memory-intensive workloads, as well as a programming interface. Coroutine switching frequency and per-switch overhead are decreased.

\item We co-design the compiler with hardware featuring decoupled memory access and dynamic scheduling to further enhance scalability and performance.

\item We propose a novel arrival-aware branch prediction mechanism that eliminates the performance loss caused by the randomness of dynamic coroutine scheduling.

\item We integrate our compiler and implement RTL of enhanced hardware into XiangShan\cite{xu2022micro}, the open-source high-performance RISC-V processor. Evaluated on memory-intensive applications, CoroAMU demonstrates average performance speedups of 3.39$\times$ and 4.87$\times$ under memory latency of 200ns and 800ns respectively.

\end{itemize}

\section{Background and Motivation} \label{sec:background}

This section introduces two variants of memory access operations—prefetching and decoupled memory access instructions—and their applications in coroutines to hide memory latency. 

\subsection{Software Prefetching}

Prefetching is a technique that proactively loads data from memory into the cache to reduce access latency and CPU stall time during actual memory operations.\cite{intro_prefetch_04icde}\cite{mittal2016acm} 
Decoupled Access and Execute (DAE)\cite{daedal} is a compilation approach that leverages software prefetching to reorganize the loops by splitting each loop iteration into multiple execution phases by memory accesses. When one iteration encounters a memory operation, it issues a prefetch instruction before switching to another iteration. However, due to register pressure and limited program semantic information, compilers can only perform conservative code transformations. Asynchronous Memory Access Chaining (AMAC)\cite{amac} employs handwritten state machines to track iteration progress and custom data structures to maintain per-iteration private data. While this manual optimization achieves high performance, it imposes significant burdens on programmers. 

\begin{figure}[hbpt]
    \centering
    \includegraphics[width=0.45\textwidth]{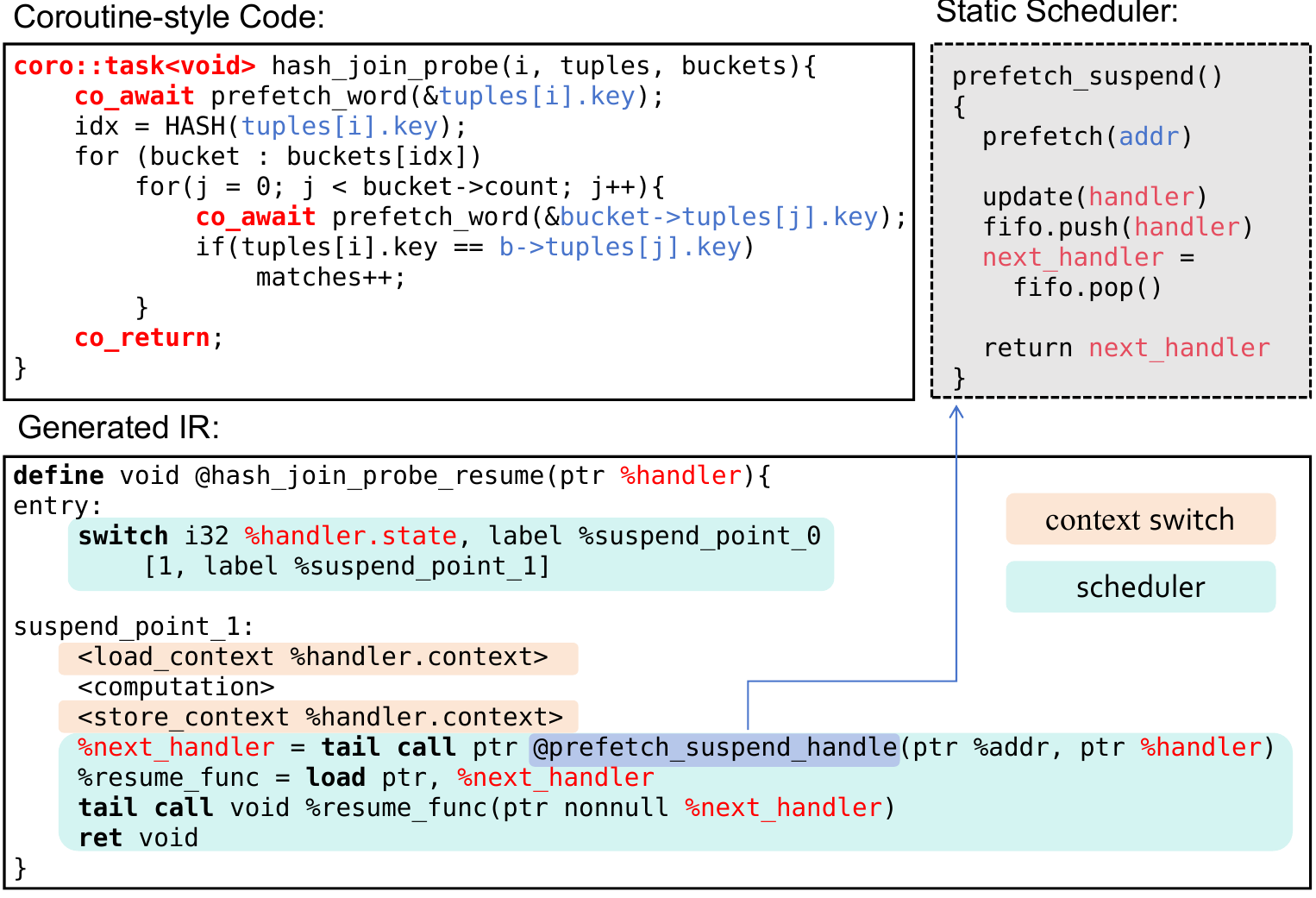}
    \caption{Coroutine usage based on software prefetching and its generated IR.}
    \label{fig:background_coro}
\end{figure}

\subsection{Coroutines}

Coroutines\cite{c++20coroutine}, introduced as a programming interface in C++20, extend traditional functions by enabling suspension and resumption through \texttt{co\_yield()} and \texttt{co\_await()} statements, with execution ultimately terminated via \texttt{co\_return()}. They delegate scheduling to a custom scheduler, allowing programmers to focus on task content rather than low-level concerns like context switching or manual optimization. By generating user-mode scheduling code, coroutines excel at masking long-latency operations by overlapping them with computational parts. Coroutines in C++20 are characterized as stackless, meaning that each task does not maintain an individual active stack. Instead, all private data is stored in a designated heap area, and the data is restored from this region when the task resumes execution. These constitute the context-switching mechanism. Unlike processes in OS, coroutines do not require saving all architectural registers; instead, the software preserves only the necessary data on demand. 

Previous works have explored memory-driven coroutines—i.e., using coroutines to hide memory access latency—applying them in database\cite{corobase} and graph computing systems\cite{corograph}, demonstrating their significant effectiveness in alleviating CPU stalls caused by memory bottlenecks\cite{prefetch_sw_coro}. Cimple\cite{decouple_cimple} presents a DSL for loop-based coroutines. The basic schema and generated code are shown in Fig.\ref{fig:background_coro}, suffering from several key limitations.

First, they all rely on software prefetching and employ a static scheduler, forcing coroutines to execute tasks in a fixed order. This rigid design prevents the code from adapting to emerging hardware and disaggregated systems.

Second, while coroutines effectively mask memory latency, coroutine switching itself becomes a new performance hotspot. The control overhead has received insufficient attention.

Third, these solutions still require experienced programmers to write efficient C++ coroutine code and are limited to simple data structures. Some even redesign execution engines to accommodate coroutines, making broader adoption challenging. A more accessible programming interface is needed to support wider applications.

\begin{figure}[hbpt]
    \vspace{-1em}
    \includegraphics[width=0.45\textwidth]{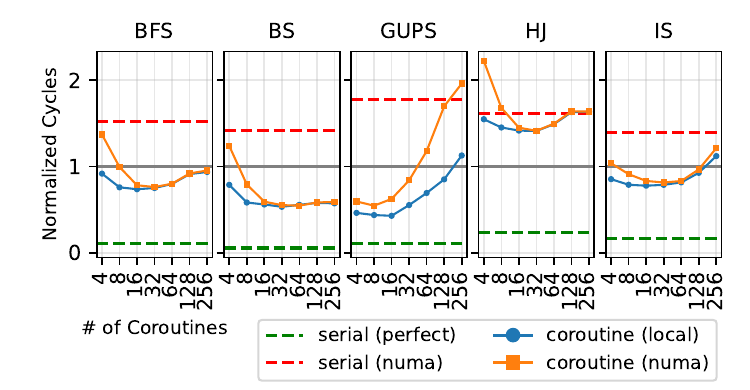}
    \vspace{-0.5em}
    \caption{Performance comparisons of serial and coroutine-based execution, normalized to serial. The red line shows the baseline performance with cross-NUMA memory accesses, and the green one shows the performance with a perfect cache.}
    \label{fig:motivation_1}
\end{figure}

\begin{figure}[hbpt]
    \centering
    \vspace{-1em}
    \includegraphics[width=0.45\textwidth]{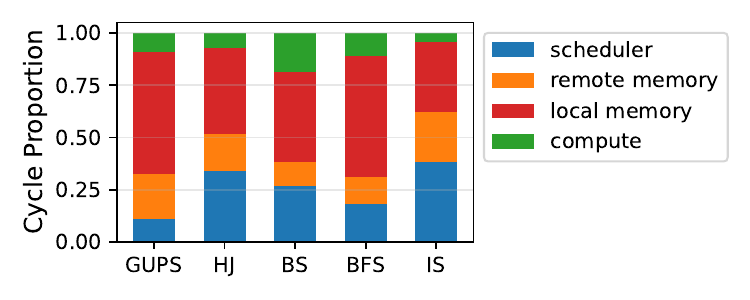}
    \vspace{-1em}
    \caption{Performance breakdown of coroutine-optimized applications. The local memory part includes context-switching overhead.}
    \vspace{-1em}
    \label{fig:motivation_2}
\end{figure}

We implement coroutines based on software prefetching for multiple benchmarks and experiment on an Intel Xeon Gold 6130 server based on the Skylake architecture, with memory accesses configured as either \textit{local} (within the same NUMA node) or \textit{numa} (crossing NUMA boundaries), demonstrating access latencies of approximately 90 nanoseconds and 130 nanoseconds respectively. Fig.\ref{fig:motivation_1} demonstrates the overall performance. While we confirm that coroutines significantly reduce the last-level cache (LLC) miss rate, the achieved performance still shows a considerable gap compared to the zero-overhead perfect prefetching.
Fig.\ref{fig:motivation_2} elucidates the runtime overhead origins, where accesses to large datasets are identified as 'remote'. The coroutine scheduler and context-switching process contribute to the most of total execution time, both accounting for more than 30\% on average. In the GUPS application, its control flow is simple, allowing the compiler to identify opportunities for optimizing the scheduler with SIMD instructions. 
The fundamental inefficiency of coroutines stems from their pursuit of generality. Conventional coroutine frameworks must accommodate heterogeneous tasks and complex scheduling demands, thereby shifting optimization burdens entirely to compilers. By exploiting memory parallelism and data locality, memory-centric coroutines can relax these constraints through simplified switching between homogeneous tasks and inter-task context sharing.

\textbf{Insight \#1}: \textit{Memory-centric coroutines unlock new possibilities for low-overhead context switching and reduced context preservation requirements.}

Moreover, the primary challenge in effectively utilizing prefetching stems from its stateless nature. Once a prefetch request is issued, the program cannot easily verify whether the data has arrived in the cache or been evicted by newer entries, potentially leading to incorrect decisions. As illustrated in Fig.\ref{fig:motivation_1}, overall performance degrades with increasing latency, and the window for the optimal number of coroutines required to achieve peak performance narrows. Additionally, this number shows inconsistency across different applications. This limitation arises because while prefetch requests avoid CPU stalls, they still consume critical MSHR entries in the cache hierarchy, and are susceptible to resource contention and cache-line conflicts. 

\textbf{Insight \#2}: \textit{To improve scalability, coroutines need to be integrated with stateful hardware to implement adaptive scheduling mechanisms.}

\begin{figure}[hbpt]
    \centering
    \includegraphics[width=0.45\textwidth]{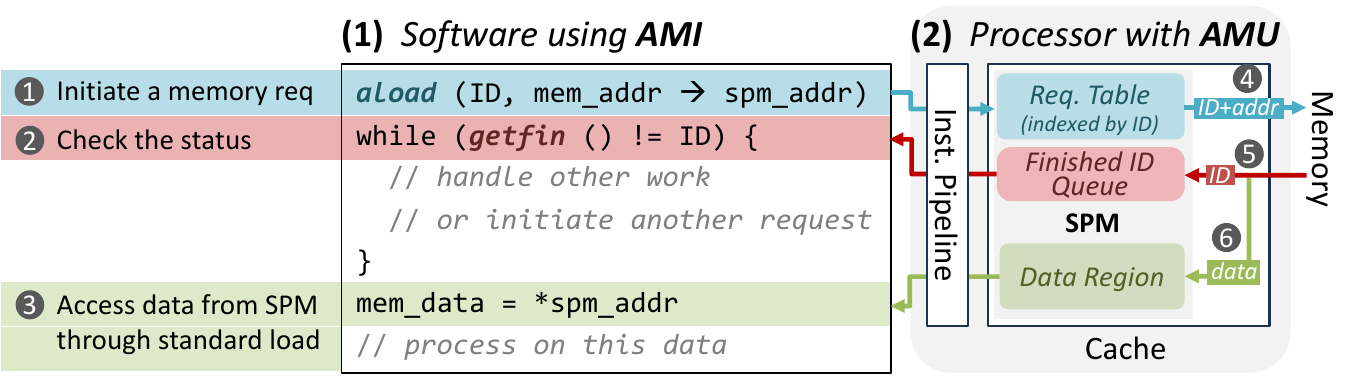}
    \caption{Basic concept of AMU.}
    \label{fig:background_amu}
\end{figure}

\subsection{Decoupled Memory Access Instructions}

Recent studies have proposed decoupled memory access techniques that provide two distinct interfaces for each memory operation: one to \textbf{issue} a memory request and another to \textbf{poll} the status of in-flight requests. This interface naturally aligns with multi-task schedulers, enabling software to determine whether to resume a suspended instruction stream based on polling feedback.

SWOOP\cite{decouple_swoop} leverages prefetch instructions as an issue interface and introduces a hardware mechanism called \texttt{Chkmiss}, which can verify whether a batch of previously issued operations has completed and arrived in the cache, thereby guiding execution flow. It uses a bitmap in the TLB entries to indicate the presence of a last-level cache line, updated upon eviction. It can only maintain a few \texttt{Chkmiss}es and corresponding memory objects at the same time.

Asynchronous Memory Unit (AMU)\cite{amu} further advances this concept with asynchronous memory instructions (AMI), introducing ID-tagged \texttt{aload} and \texttt{astore} operations and a \texttt{getfin} instruction to retrieve one of the IDs of completed requests. (See Fig.\ref{fig:background_amu}) To sustain high memory-level parallelism (MLP), AMU dedicates a scratchpad memory (SPM) region within the private L2 cache to store both metadata and data for all in-flight requests. (1) Users can interact with it using standard load/store instructions for direct data manipulation, and enable asynchronous, non-blocking data movement between the SPM and main memory via \texttt{aload} and \texttt{astore} instructions. (2) A portion of the SPM is reserved as a hardware-managed Request Table, functionally analogous to Miss Status Handling Registers (MSHR) in traditional cache systems. An entry is allocated to track request details, including memory address, request ID, data size, and status.

Recent work has focused on on-chip accelerators designed to offload frequent but fine-grained memory-intensive operations, such as serialization and deserialization in server workloads\cite{intro_acc_21micro}. These accelerator controllers expose a control interface to the CPU, adhering to the same issue-poll paradigm.

However, employing dynamic schedulers incurs non-trivial tradeoffs. Frequent and stochastic memory events create control flow divergence that resists static analysis, while the control transfers between coroutines manifest as unpredictable branches. In modern architectures, mispredicted branches could incur tens of cycles of pipeline bubbles\cite{motivation_branch}, ultimately compromising efficiency.

\textbf{Insight \#3}: \textit{The CPU frontend requires co-design with decoupled memory access modules to enable efficient coroutine switching.}

Hardware employing decoupled memory access also lacks mature compiler optimization (e.g., vectorization and group prefetching), resulting in excessive coroutine switching demands and suboptimal performance that sometimes fails to match static scheduling approaches.

\textbf{Insight \#4}: \textit{To minimize switching frequency, it would benefit if the hardware supports coalescing multiple memory operations for the compiler to exploit.}

Based on these insights, we present CoroAMU, a hardware-software co-design framework that contains a compiler tool that generates memory access coroutines based on LLVM and an enhanced memory engine design based on AMU.
While this work adopts AMU as the basis, it is worth noting that the design principles generalize to any future hardware supporting issue-poll interfaces.

\begin{figure}[hbpt]
    \centering
    \includegraphics[width=0.45\textwidth]{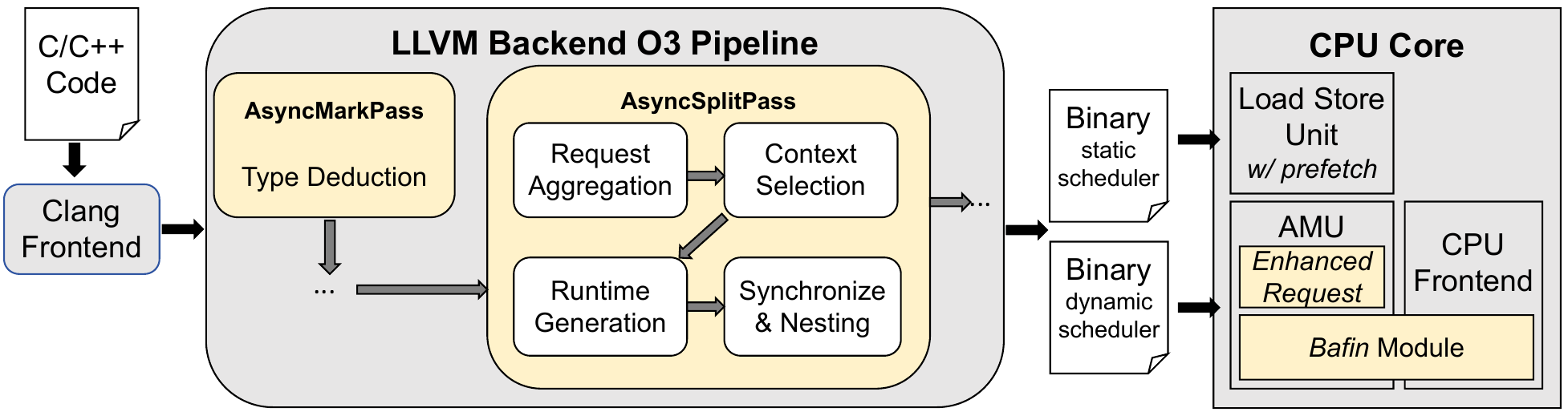}
    \caption{CoroAMU overview, with a compiler automatically generating coroutines and enhanced AMU hardware.}
    \label{fig:design_overview}
\end{figure}

\section{Software Components}

Fig.\ref{fig:design_overview} shows the overview of CoroAMU. In this section, code generation by CoroAMU from the original C/C++ source code is covered.
Sections \ref{sec:soft_prog} and \ref{sec:soft_ctx} analyze redundancies in runtime and context switching generated by current coroutines and our optimizations. Section \ref{sec:soft_agg} introduces our approach to reducing switch frequency by leveraging memory access patterns. Section \ref{sec:soft_bafin} addresses the performance overhead of dynamic scheduling, particularly extra branch misprediction. Sections \ref{sec:soft_sync} and \ref{sec:soft_nest} describe how our programming model supports synchronization and nested coroutines to ensure correctness and compatibility. Finally, we detail the compiler implementation and its interfaces.

\subsection{Programming Model} \label{sec:soft_prog}

CoroAMU specifically targets all memory-intensive \texttt{for} loops without requiring them to be canonical or innermost loops imposed by prior DAE work\cite{clairvoyance}. This design enables seamless integration with most OpenMP \texttt{for} loops with minimal modifications. 

Unlike C++20's interface which demands manual implementation of suspension points, schedulers, coroutine allocation and construction, and barriers, CoroAMU automatically generates complete runtime code encompassing all these components. Each iteration is treated as a complete coroutine task, thus all of the tasks share homogeneous code sections,
allowing us to consolidate runtime and actual tasks within a single function. 
This design avoids challenging inter-procedural optimizations and possible overhead of stack management.

\begin{figure}[hbpt]
    \centering
    \includegraphics[width=0.4\textwidth]{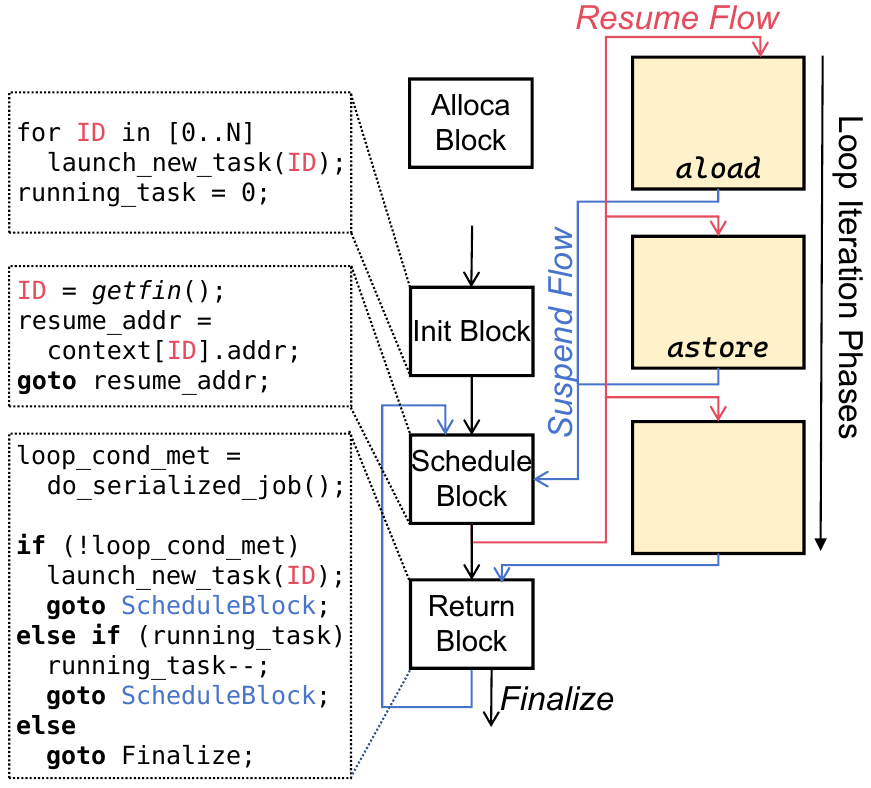}
    \caption{Basic code generation of CoroAMU.}
    \label{fig:design_program}
\end{figure}

Basic generated code comprises several components (Fig. \ref{fig:design_program}):

\textbf{Alloca Block}: establishes a handler array to maintain execution states of active iterations while allocating fixed-size storage for coroutine context. Memory requirements are determined at compile-time to eliminate dynamic allocation overhead.

\textbf{Init Block}: functions as the primary entry point, initializing all metadata structures and launching the initial coroutine batch.

\textbf{Schedule Block}: implements the core scheduling logic. For static prefetch scenarios, it adopts a FIFO queue to select the next ready coroutine. In dynamic cases, it uses hardware polling mechanisms to retrieve a handler.

\textbf{Return Block}: serves as the lifecycle manager. It handles coroutine termination by recycling handlers, initializing subsequent iterations, and spawning new coroutines. It also sequentially handles loop-carried dependencies. Under dynamic scheduling, this block periodically adjusts concurrency levels based on polling feedback to adapt to memory latency.

\textbf{Loop Phases}: preserves parallel segments, issuing non-blocking prefetch or AMU requests before long-latency operations and voluntarily yielding CPU control.

\subsection{Optimizing Context} \label{sec:soft_ctx}

In stackless coroutine implementations, the runtime allocates dedicated memory space to store intermediate computational results during execution. It is performed through static analysis of each static single assignment (SSA) variable's definition-use chain. By tracking variable lifetimes across suspension points, it identifies which values must be saved in the coroutine context.

When all coroutines originate from the same loop iteration, the context constructed contains redundant elements. 
The most straightforward case involves read-only variables shared across iterations. They can safely bypass context analysis and be accessed directly. The remaining ones can be categorized into three types based on how they are updated within coroutines: (1) Variables whose updates depend solely on their immediate context rather than their previous value are coroutine-private and must be saved in the context; (2) Some variables are updated depending on previous values but exhibiting commutative behavior, which means the update order does not affect final results. They can be accessed in a shared manner just like read-only variables; (3) In complex or ambiguous cases, the compiler conservatively extracts such variables for serialized updates before coroutine launch or after completion.

The variables are thus classified into three categories: private, shared, and sequential. While the compiler performs static analysis for scalar variables, precise classification requires hints provided by programmers. 

\subsection{Coalescing Memory Requests} \label{sec:soft_agg}

Long-latency memory accesses occur frequently in memory-intensive applications, each triggering a coroutine switch that occupies the effective execution time of actual workloads. An effective solution is to issue multiple memory requests at a time before the coroutine yields. However, this approach has rarely been explored in coroutines because I/O events are typically infrequent, and it
may introduce deadlock risks.


Achieving aggregation in the issue/poll mode requires careful design. Stateful memory access requests need to explicitly bind to a handler upon issuance (an ID in AMU), and polling involves querying the status of all requests bound to this handler. Two scenarios are identified where requests can be merged within a single coroutine. The first case is coarse-grained requests. When memory operations are performed in the same memory region, such as fields of a data structure, it can fetch or write a batch of data (up to 4KB) in one request. The second one involves requests without data dependencies. Two requests are independent if neither depends on the memory result of the other, safe to be issued concurrently.

The hardware interface is modified to accommodate these two requirements. CoroAMU extended the high-order bits of the memory address field in \texttt{aload}/\texttt{astore} instructions to encode granularity. For independent requests, we introduced the \texttt{aset} instruction, which takes two operands: a handler ID and a positive integer \texttt{n}, indicating that the next \texttt{n} requests are bound to this ID. The CPU can only retrieve this ID via \texttt{getfin} once all these accesses are completed. The architectural modifications will be detailed in section \ref{sec:arch}.

During instruction scheduling, some general constraints must be satisfied. When trying to schedule two instructions together, the process must preserve existing data dependencies, memory consistency, and side-effect barriers. Furthermore, the number of aggregated requests must not exceed the hardware capability. This formulates a constrained instruction scheduling problem with an optimization objective: minimizing context switches after request aggregation. We observe that it is sufficient to find merging candidates in a basic block. Therefore a simple greedy algorithm inside each basic block is implemented. 

\subsection{Enhanced Dynamic Scheduler} \label{sec:soft_bafin}

\begin{figure}[hbpt]
    \includegraphics[width=0.48\textwidth]{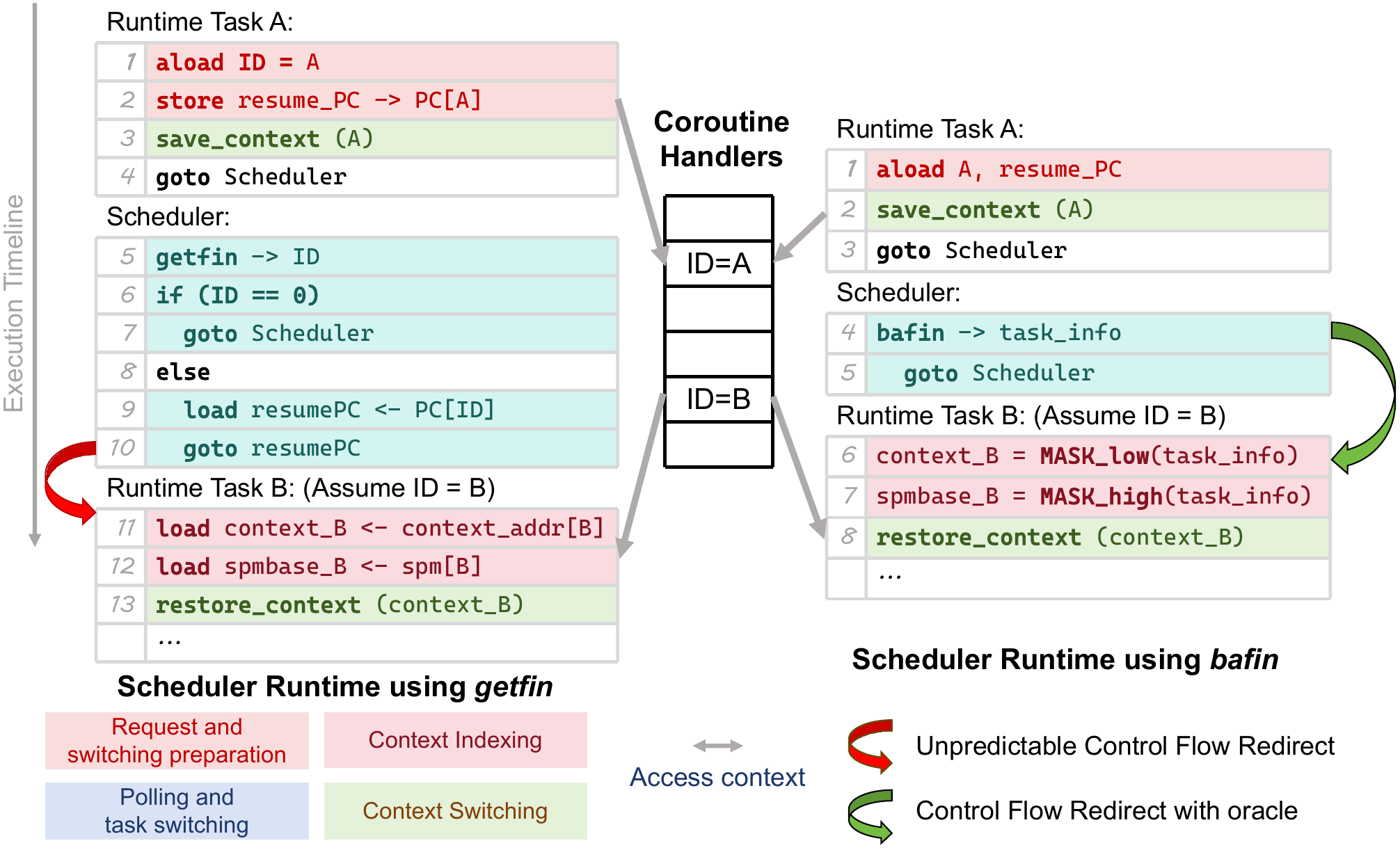}
    \caption{Scheduler comparison between \textit{getfin} and \textit{bafin}.}
    \label{fig:design_bafin}
\end{figure}

The dynamic scheduling of coroutines introduces an unpredictable indirect jump, stemming from the scheduler's need to poll before deciding which coroutine to resume. 
As shown in Fig. \ref{fig:design_bafin} (left), when task A initiates a coroutine switch due to an asynchronous memory access, it first stores the PC of the resume address in the metadata area. The scheduler then suspends task A, selects a coroutine that has completed memory access, and redirects the execution flow through an indirect jump.
As the memory operations increase, so does the number of suspension points and potential jump targets, making the indirect jump effectively random.

The fundamental idea for eliminating misprediction is as follows: when a memory access request is initiated, the resume PC corresponding to this suspension is bound to the request. Once the access is completed, this jump target can be directly transmitted from the memory unit to the branch prediction unit (BPU) in the CPU frontend.
A new jump instruction is required to signal the BPU to utilize the guiding information for prediction. Original AMU used \texttt{getfin} instructions to retrieve the ID of a completed memory access. It is replaced by \texttt{bafin} instruction, which performs a jump directly to the coroutine resumption point based on a completed ID or falls through if no such ID exists. The jump target corresponding to each ID is encoded in the high-order bits of the address in the \texttt{aload}/\texttt{astore} requests. 

The \texttt{bafin} instruction provides additional benefits for schedulers by eliminating the need to maintain the jump targets or status of coroutines. This insight suggests that other ID-indexed metadata, such as context address, can also be offloaded to hardware. To achieve this, a custom \texttt{aconfig} instruction is used to configure two dedicated hardware registers: one stores the base address of the handler array, and the other specifies the handler size per coroutine. When \texttt{bafin} is executed, the hardware calculates its corresponding handler address and writes it back to a register operand. As illustrated in Fig. \ref{fig:design_bafin}, using \texttt{bafin} for coroutine switching allows the scheduler to perform just two easy-to-predict jumps and three ALU operations.

\begin{figure}[hbpt]
    \centering
    \includegraphics[width=0.48\textwidth]{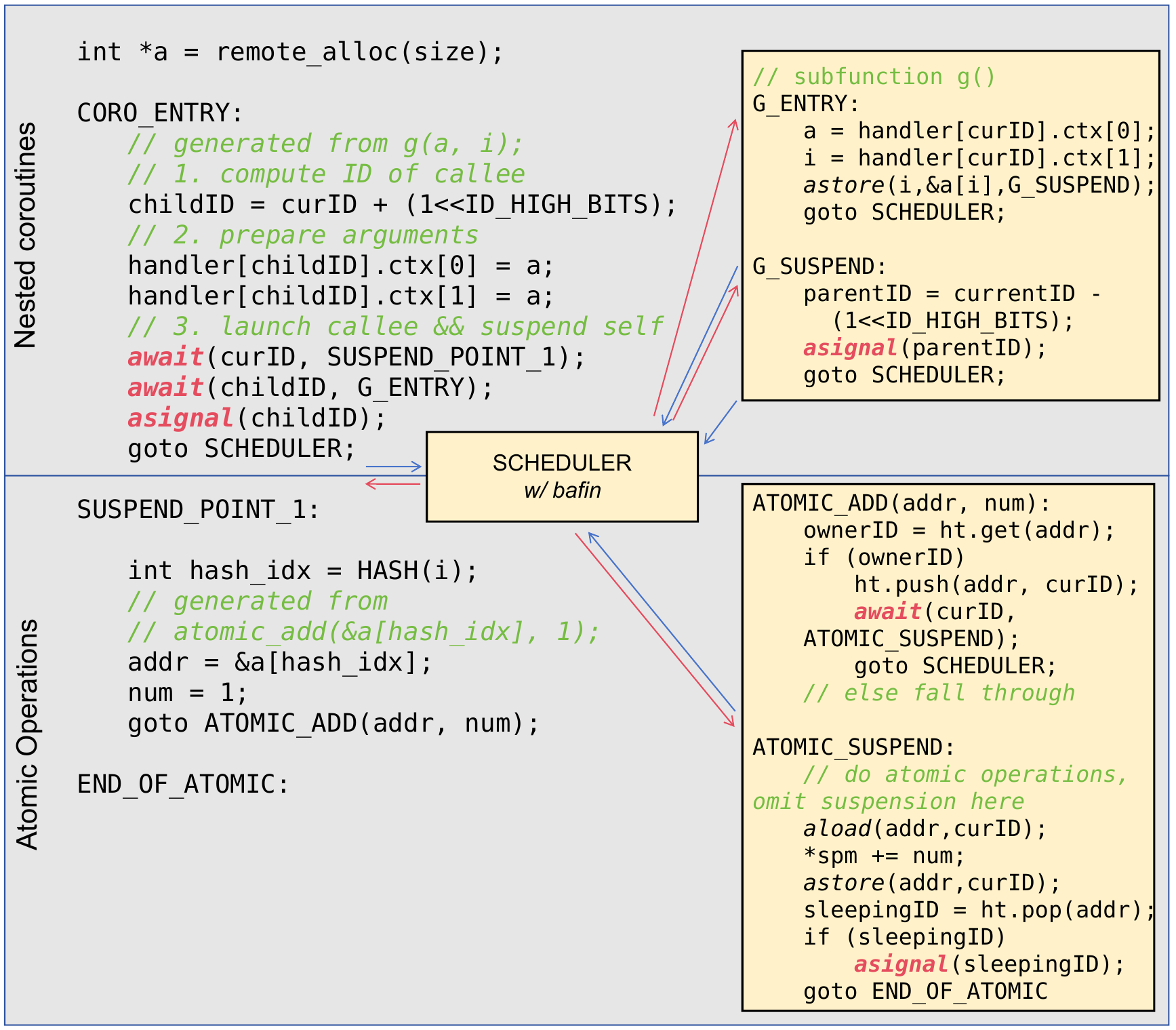}
    \caption{Synchronization and nested coroutine supported by \textit{await} and \textit{asignal}.}
    \label{fig:design_sync}
\end{figure}

\subsection{Memory Consistency and Synchronization} \label{sec:soft_sync}

Similar to multi-threading, transforming a loop into pausable coroutines brings memory consistency concerns. CoroAMU faces two-fold consistency: traditional core-to-core consistency, and consistency between coroutines within a core. Traditional consistency issues arise when multiple threads simultaneously read and write the same memory locations. In contrast, coroutines face consistency issues only if suspension points split critical sections, which is less problematic.
For example, reduction variables typically do not require protection since they undergo simple read-modify-write operations that remain uninterrupted by coroutine suspension.

In multi-core systems, programmers employ locks or atomic instructions to prevent inconsistencies. Atomic instructions are reused in CoroAMU, which serve two primary purposes: (1) The operation object of the atomic instruction itself is a long-latency memory object, whose delay needs to be masked; and (2) atomic instructions can construct locks to protect critical sections. Our compiler automatically identifies and transforms these atomic instructions into a software procedure. The example of atomic operations is illustrated in Fig.\ref{fig:design_sync}. The coroutine with \texttt{curID} attempts to apply atomic addition to \texttt{a[hash\_idx]}. First, it queries a hash table (\texttt{ht}) indexed by the object memory address.
If another coroutine is holding the ownership, the current one suspends itself by pushing itself into the hash table and yields execution until explicitly waken up. Upon resumption, it proceeds with the atomic operation before finally popping and signaling the next waiting coroutine.

CoroAMU provides a dedicated interface through \texttt{await} and \texttt{asignal} instructions. When a coroutine needs to suspend itself and avoid being scheduled, it invokes the \texttt{await} instruction, which registers its handler. Conversely, the coroutine releasing resources executes the \texttt{asignal} instruction, effectively responding to the prior \texttt{await} request with the specified ID. This signaling mechanism ensures the ID becomes visible to subsequent \texttt{getfin} or \texttt{bafin} operations.

\subsection{Nested Coroutine} \label{sec:soft_nest}

Internal function calls bring challenges for coroutine metadata management.
If called functions do not contain long-latency memory operations, they can be executed as normal without disrupting coroutine integrity. For callees requiring suspension, most of them are inlined
as the simplest solution. In cases where inlining is not worthwhile, CoroAMU extends the C++ nested coroutines. Traditional C++ coroutines require programmers to explicitly wrap child functions into coroutine tasks
and use \texttt{co\_await()} for control transfer. This approach complicates schedulers and demands programmer intervention.

Our solution leverages the \texttt{await}/\texttt{asignal} instructions as in Fig.\ref{fig:design_sync}. Before invoking a child function, the caller (with \texttt{curID}) executes \texttt{await} to suspend itself while assigning the child a derived ID by incrementing the high-order bits. The child is initialized via a pair of \texttt{await} and \texttt{asignal} instructions: the former registers the child's handler into the SPM, while the latter enables the childID to be acquired by the scheduler in a ready state. Upon the child's return, it computes the caller's ID and resumes the caller via \texttt{asignal}.

\lstset{language=C, escapeinside=``, caption=Example usage of compiler interface.,label=code:compiler_interface}
\begin{lstlisting}  
// Tuples and buckets are large data structures may stored
// in the disaggregated memory, require transformation.
#pragma asyncmem num_task(`\textcolor{red}{\texttt{64}}`) shared_var(`\textcolor{green!35!black}{\texttt{matches}}`)
for (i = 0; i < num_tuples; i++) {
    __builtin_is_remote(`\textcolor{red}{\texttt{ht->buckets}}`);
    __builtin_is_remote(`\textcolor{red}{\texttt{tuples}}`);
    bucket_t *b = ht->buckets + HASH(`\textcolor{red}{\texttt{tuples[i].key}}`);
    while(b) {
        for (j = 0; j < `\textcolor{red}{\texttt{b->count}}`; j++)
            if (`\textcolor{red}{\texttt{tuples[i].key}}` == `\textcolor{red}{\texttt{tuples[j].key}}`) `\textcolor{green!35!black}{\texttt{matches}}`++;
        b = `\textcolor{red}{\texttt{b->next}}`;
    }
}
\end{lstlisting}

\subsection{Implementation and API}

We implemented our compiler on Clang/LLVM infrastructure. The compiler processes C/C++ programs where memory-intensive \texttt{for} loops are annotated with \texttt{pragma}s, extending the LLVM pass pipeline with two passes: AsyncMarkPass and AsyncSplitPass. During compilation, the system generates either a static scheduler using \texttt{prefetch} instructions or a dynamic scheduler leveraging \texttt{getfin}/\texttt{bafin} instructions, capable of running on both unmodified CPUs and those supporting AMU.

As shown in Listing \ref{code:compiler_interface}, \texttt{pragma}s are used to identify loops for coroutine transformation and gather essential compiler directives, including suggested parallelism and variable hints. CoroAMU employs distinct address spaces to distinguish remote regions, with accesses to these areas transformed into coroutine suspension. It is assumed that each pointer's characteristics remain static throughout execution. This allows us to leverage LLVM's address space attribute for strict typing, a mechanism traditionally used for GPU memory. Remote memory is explicitly allocated with the \texttt{remote\_alloc()} interface, or annotated with custom \texttt{built-in}s.

CoroAMU adopts two-phase IR passes. The first phase executes immediately after IR generation, while the second phase follows most \texttt{O3} optimization passes. It serves dual purposes: (1) allowing conventional optimizations to eliminate redundant variables, thereby minimizing context, and (2) postponing control flow modifications that could otherwise limit optimization opportunities around suspension points. AsyncMarkPass operates transparently by annotating variables and memory operations, maintaining full optimization potential.

\section{Architectural Support with AMU} \label{sec:arch}

\begin{figure}[hbpt]
    \centering
    \includegraphics[width=0.45\textwidth]{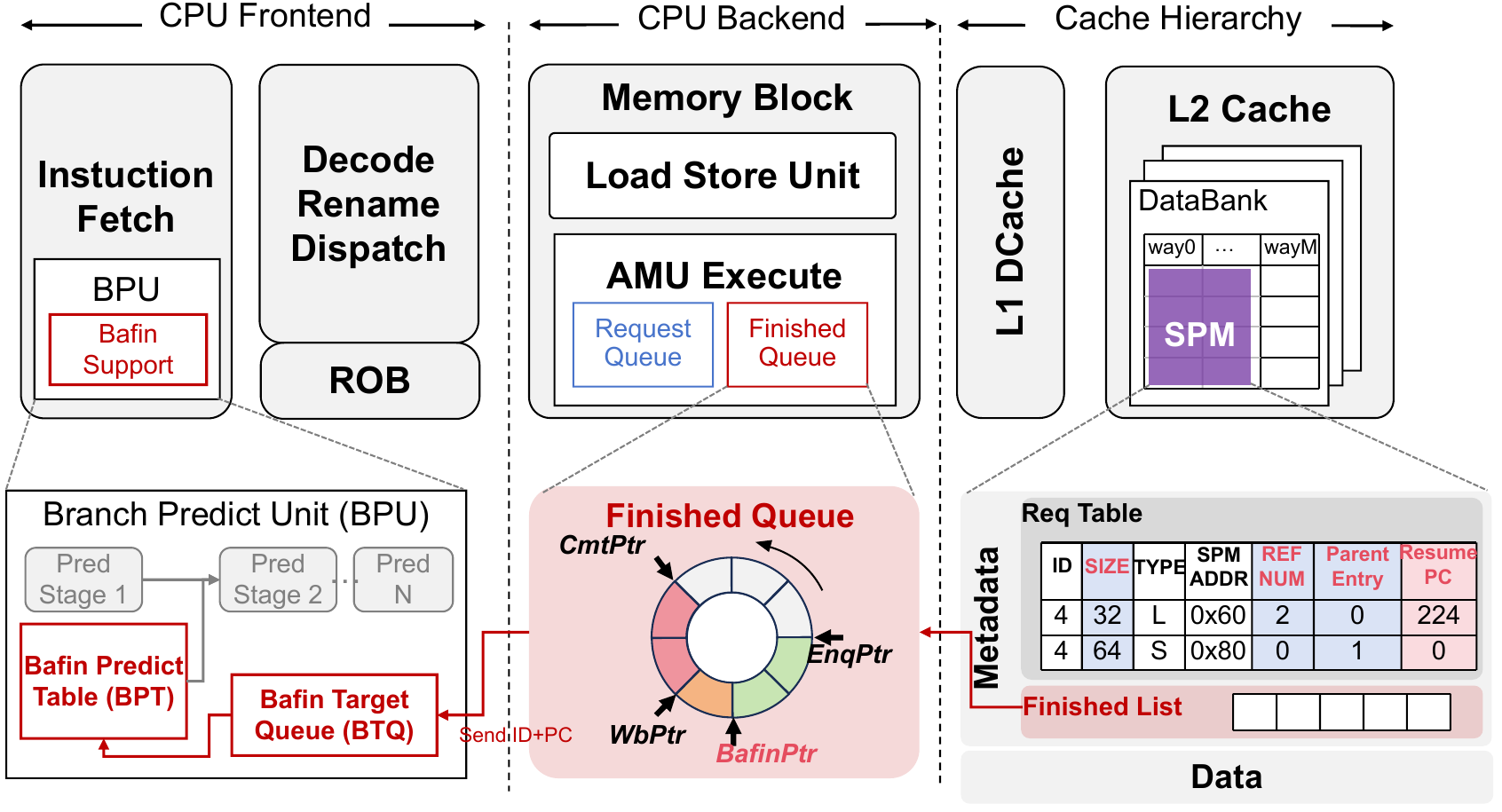}
    \caption{Overview of hardware modifications.}
    \label{fig:design_arch}
\end{figure}

Fig.\ref{fig:design_arch} illustrates the architectural extension of CoroAMU built upon the existing AMU hardware to enhance dynamic scheduling performance for coroutines. The extension introduces three key sets of novel instructions and software interfaces: \texttt{bafin} that polls and jumps to coroutines while eliminating branch prediction penalties (colored red), \texttt{aset} and enhanced \texttt{aload}/\texttt{astore} to aggregate memory access of variable granularity, and \texttt{await}/\texttt{asignal} that enable proactive coroutine sleep and wake-up operations (both colored blue).

\subsection{Eliminating Branch Misprediction}

The key machenism of eliminating branch misprediction is having the memory unit provide completed memory access and its resume target to the front-end BPU. The \texttt{bafin} instruction is predicted like conventional branches, and it consumes this oracle to guarantee accurate prediction. This mechanism requires three primary hardware modifications: the Bafin Prediction Table (BPT) in the BPU for oracle consumption, the Finished Queue in the memory subsystem for oracle generation, and the Bafin Target Queue (BTQ) serving as the communication bridge.

The Bafin Predict Table is a specialized predictor similar to an L1 BTB but tracks only \texttt{bafin} instructions, indexed by PC. Designed with minimal storage (4 entries in our implementation), it efficiently serves its purpose since \texttt{bafin} only appears in scheduler code and our software model maintains just one active scheduler. Operating with single-cycle latency, it enables bubble-free rapid prediction by retrieving an ID and corresponding jump address from the BTQ as a result. BPT has a higher priority than other predictors.

The Finished Queue in AMU is an execution unit for polling instructions and stores IDs of completed memory requests. An additional field \texttt{PC} is added to the queue entries, as well as to the request entry in the L2 cache, to maintain the offset of the resumption target. When an \texttt{aload}/\texttt{astore} instruction is executed, it carries the target offset in the high-order bits of its address along to the L2 cache. Upon arrival of the response, both the ID and corresponding PC offset are transferred to the Finished Queue. The IDs available for delivery are specifically those that have entered the Finished Queue but have not yet been written back by \texttt{getfin}/\texttt{bafin} instructions.

The Bafin Target Queue reliably delivers IDs during processor redirection events. It implements two distinct rollback mechanisms. When the pre-decode unit detects a misprediction, the BTQ efficiently reverts the IDs after the misprediction into an unused state, preserving the contents of the queue. For back-end redirections, the BTQ flushes all queued IDs and requests the Finished Queue to resend them.

\subsection{Aggregated Memory Request}

CoroAMU implements two distinct memory access aggregation mechanisms to reduce switching frequency. 
The first approach employs enhanced \texttt{aload}/\texttt{astore} instructions capable of accessing multiple contiguous words in a single operation. The second utilizes the \texttt{aset} instruction to coalesce subsequent requests. Both methods share identical hardware logic.

At the execution stage, these coarse-grained instructions undergo decomposition into cache-line-sized requests. The L2 cache receives a batch of requests sharing an identical ID, with the first arriving request designated as the primary request. Subsequent requests spawn new entries in the Request Table while maintaining references to their primary request. A counter is implemented for each group, incremented upon receiving responses for constituent operations. Only when all expected responses are collected does the memory unit generate a single completion notification to the CPU.

\subsection{Waiting and Signaling}

The \texttt{await} and \texttt{asignal} instructions form a hardware primitive for synchronization and nested coroutines. These simple yet powerful instructions reuse existing AMU components. The \texttt{await} instruction functions as a non-access \texttt{aload} operation that simply registers its ID in the Request Table without generating actual memory traffic. Conversely, the \texttt{asignal} instruction acts as a response that matches a pending request, subsequently pushing the ID into the Finished Queue. In essence, this method uses the Request Table and the Finished Queue, to distinguish whether a coroutine is hung-on or ready-to-run, offloading the work of the software scheduler.

\section{Experimental Setup} \label{sec:experiment}
The compiler experiments were conducted on a two-socket Intel Xeon Gold 6130 server with 128GB DDR4 of 2666 MT/s.
CoroAMU hardware is implemented with Chisel HDL on \textsc{Nanhu}, the second generation microarchitecture of XiangShan\cite{xu2022micro}, which is currently the open-source RISC-V processor with the highest performance and also industry-competitive\footnote{The normalized performance of Nanhu is reported as 7.94/GHz on SPECint2006. Competitors: XT-910 achieves 6.11/GHz, Cortex-A73 achieves 6.75/GHz.}.

\begin{figure}[hbpt]
    \centering
    \includegraphics[width=0.45\textwidth]{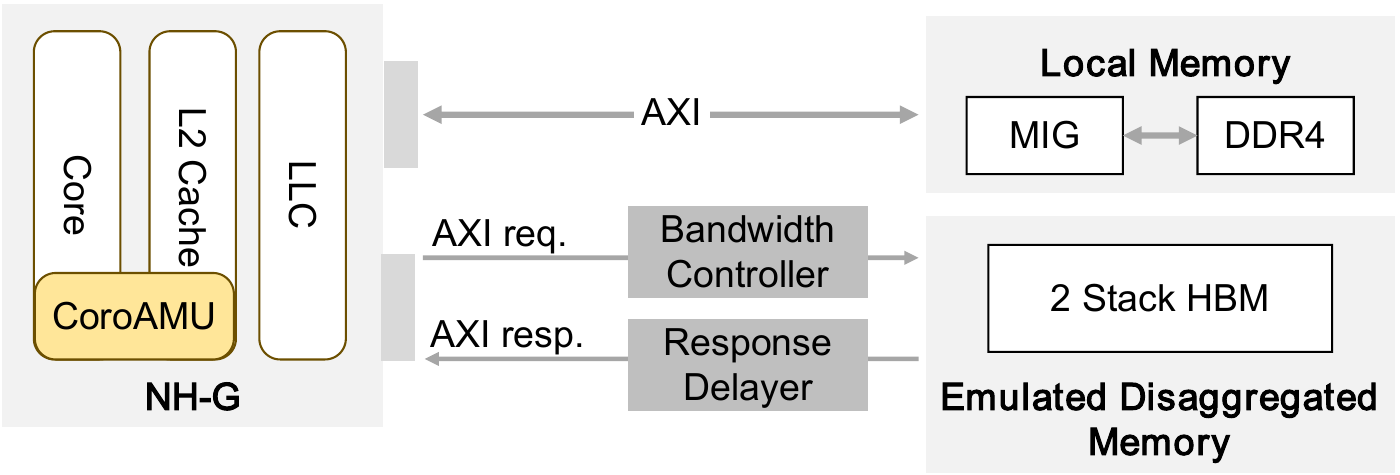}
    \caption{Overview of the FPGA Prototype.}
    \label{fig:fpga_proto}
\end{figure}

\textbf{Evaluation Environment Setup on FPGA} 
We implement a prototype emulating a disaggregated memory system on Xilinx VCU128 FPGA with two memory subsystems (Fig. \ref{fig:fpga_proto}):
Local memory leveraging onboard DRAM, and far memory emulation via high-bandwidth memory (HBM), both interfaced through the on-chip AXI bus.
A programmable bandwidth regulator modulates far-memory request traffic while a delayer enforces configurable latency.

\textbf{Core Configuration} We developed \textsc{NH-G} (Table \ref{tab:eval_detail_nanhug}) as an FPGA-tailored variant of \textsc{Nanhu}, operating at 50MHz while maintaining architectural fidelity through proportional scaling. This prototype emulates a 3GHz processor experiencing 100ns-1us memory latency (300-3000 cycles) and 3-96GB/s bandwidth (1-32B/cycle). AMU needs to allocate a portion of L2 Cache as SPM for storing data and metadata. In our experiments, 32KB (1 out of 8 ways) is used, which is sufficient to accommodate 512 concurrent coroutines.

\begin{table}[hbpt]
\caption{Core microarchitecture configuration of NH-G.}
\begin{tabular}{|c|c|}
\hline
\textbf{Core Configuration} & \textbf{Parameter} \\
\hline
Instruction Fetch Width & 32B \\\hline
Branch Predictor & \begin{tabular}[c]{@{}c@{}}BTB + RAS +  TAGE + ITTAGE\end{tabular} \\\hline
Decode/Rename/Issue Width & 4/4/8 \\\hline
\begin{tabular}[c]{@{}c@{}}Dispatch Queue Size (Int/Fp/LS)\end{tabular} & 12/12/12 \\\hline
ROB Entries & 96 \\\hline
\begin{tabular}[c]{@{}c@{}}Int/FP Physical Registers\end{tabular} & 64/64 \\\hline
\begin{tabular}[c]{@{}c@{}}Load/Store Queue Entries \end{tabular} & 32/16 \\\hline
\begin{tabular}[c]{@{}c@{}}AMU Req/Finish Queue Entries \end{tabular} & 16/16 \\\hline
\begin{tabular}[c]{@{}c@{}}L1 D-Cache\end{tabular} & 8-way 32KB, 16 MSHRs \\\hline
L2 Cache & \begin{tabular}[c]{@{}c@{}}4 Slices, 8-way 256KB, 56 MSHRs\\ Best-Offset Prefetcher (BOP) \end{tabular} \\\hline
L3 Cache (LLC) & \begin{tabular}[c]{@{}c@{}}4 Slices, 6-way 1536KB, 56 MSHRs\end{tabular} \\
\hline
\end{tabular}
\label{tab:eval_detail_nanhug}
\end{table}

\textbf{Benchmarks} 
CoroAMU specifically targets memory-bottlenecked workloads. We aim to investigate how effectively coroutine-based single-core execution can utilize memory bandwidth. Our evaluation employs 8 benchmarks covering various memory access patterns
as listed in Table \ref{tab:benchmarks}. These applications are compiled using LLVM 18.1.0.

Experiments are conducted under Linux kernel 5.16, with memory-intensive datasets intentionally placed in the emulated far memory. Datasets are sized to exceed the capacity of the cache hierarchy. 


\begin{table}[hbpt]
\caption{Benchmarks and transformed structures.}
\begin{tabular}{|c|c|p{4cm}|}
\hline
\textbf{Suite} & \textbf{Benchmark} & \textbf{Remote Structure}\\
\hline
HPCC\cite{bench_hpcc} & GUPS & \texttt{Table}\\
\hline
Binary Search & BS & \texttt{sorted\_array}\\
\hline
Graph500\cite{bench_graph500} & BFS & graph, \texttt{bfs\_tree}, \texttt{vlist}\\
\hline
STREAM\cite{bench_stream} & STREAM & \texttt{a}, \texttt{b}, \texttt{c}\\
\hline
Hash Join\cite{bench_hj} & HJ & \texttt{relation->tuples}, \texttt{ht->buckets}\\
\hline
\multirow{2}{*}{SPEC2017\cite{spec17}} & 505.mcf\_r  & \texttt{net->nodes}, \texttt{net->arcs}\\
\cline{2-3}
 & 519.lbm\_r  & \texttt{srcGrid}, \texttt{dstGrid}\\
\cline{1-3}
NPB\cite{nas} & IS & all of \texttt{malloc()}\\
\hline
\end{tabular}
\label{tab:benchmarks}
\end{table}

\section{Evaluation}
We evaluate the following combinations of compiler and microarchitectural configurations:
\begin{itemize}
    \item \textbf{Serial}: Unmodified applications on the baseline processor.
    \item \textbf{Coroutine}: A manually written coroutine implementation with prefetch-enabled static scheduling \cite{prefetch_sw_coro}.
    \item CoroAMU with static scheduler (\textbf{CoroAMU-S}): Statically scheduled coroutines with basic code generation of our compiler.
    \item CoroAMU with dynamic scheduler (\textbf{CoroAMU-D}): An original-AMU-assisted coroutine approach (using \texttt{getfin}) with basic code generation of dynamic scheduling.
    \item CoroAMU with full optimizations (\textbf{CoroAMU-Full}): It integrates enhanced AMU (using \texttt{bafin}) with all compiler optimizations.
\end{itemize}

\subsection{Main Results}

\begin{figure}[hbpt]
    \centering
    \vspace{-0.5em}
    \includegraphics[width=0.48\textwidth]{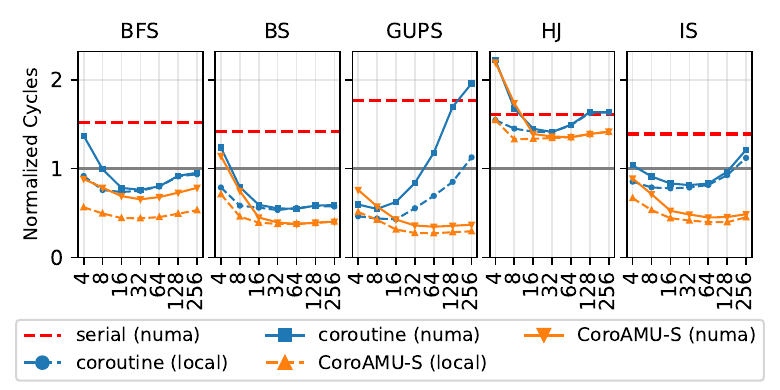}
    \caption{Performance for prefetch-based CoroAMU compiler with different number of coroutines normalized to baseline serial code on Intel Xeon Gold 6130 server.}
    \label{fig:eval_x86_prefetch}
\end{figure}

Figure \ref{fig:eval_x86_prefetch} demonstrates the performance of the CoroAMU compiler and hand-crafted coroutines on x86 servers. Under a typical latency of 100ns, SOTA coroutines generally achieve peak performance at concurrency levels between 8 and 32. Compared with serial code, they deliver average speedups of 1.40$\times$ and 2.01$\times$ for intra-NUMA and cross-NUMA respectively. CoroAMU outperforms coroutines across almost all configurations, achieving 2.11$\times$ and 2.78$\times$ average improvements with a significantly expanded optimal performance window. This enhancement stems from CoroAMU's reduced control overhead in coroutine scheduling, enabling faster instruction stream switching and more intensive prefetch generation and utilization. However, \texttt{HJ} shows limited prefetch effectiveness due to its partitioning of large datasets, and the low-latency environment prevents sufficient LLC misses to block the CPU pipeline. Additionally, the compiler inherits prefetch limitations including performance degradation at high concurrency levels and sensitivity to latency variations.

\begin{figure*}[hbpt]
    \centering
    \includegraphics[width=1\textwidth]{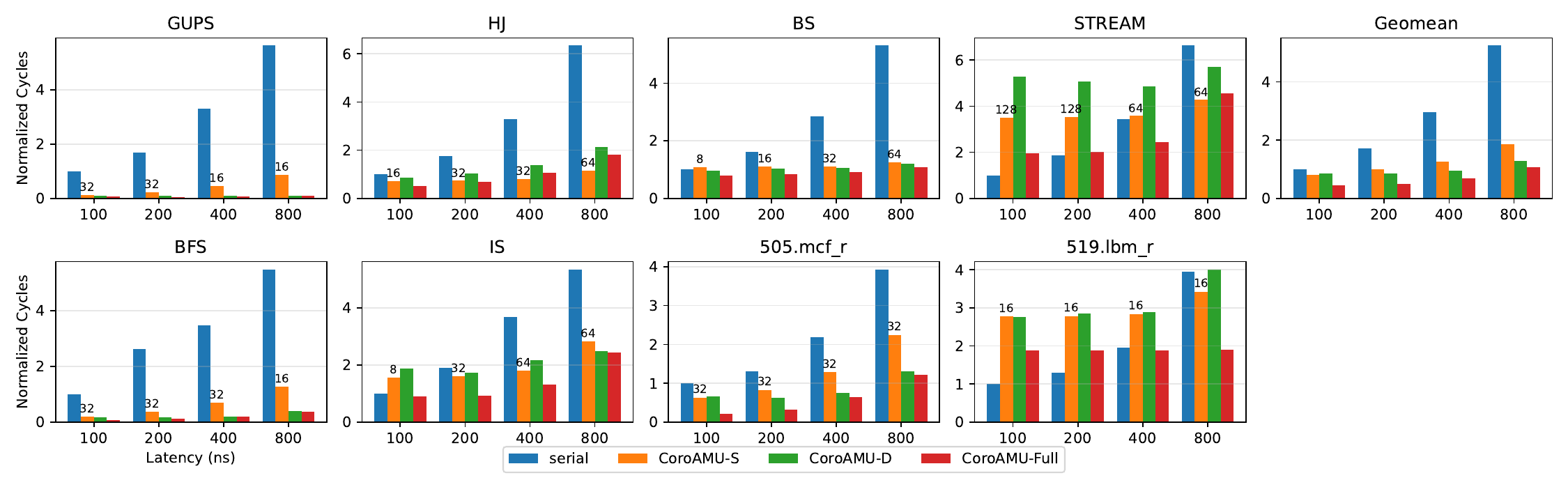}
    \caption{Performance for CoroAMU normalized to baseline serial code on NH-G. Numbers of coroutines of best performance for prefetching (CoroAMU-S) are labeled. CoroAMU-D and CoroAMU-Full are configured with 96 coroutines. }
    \label{fig:eval_main}
\end{figure*}

Figure \ref{fig:eval_main} illustrates performance with memory-access-decoupled hardware support. As memory latency increases, serial implementations exhibit near-linear runtime escalation, while CoroAMU maintains performance with marginal degradation. At 200ns and 800ns memory latencies, it achieves average speedups of 3.39$\times$ and 4.87$\times$ respectively (up to 29.0$\times$ and 59.8$\times$ in \texttt{GUPS}), demonstrating adaptability to disaggregated memory with extended latencies.

CoroAMU shows exceptional performance in \texttt{GUPS} and \texttt{BFS} workloads, which feature massive datasets, fully random memory access patterns, and minimal computation. These characteristics make them severely latency-bound. In contrast, bandwidth-bound benchmarks like \texttt{STREAM}, \texttt{IS}, and \texttt{lbm} exhibit strong spatial locality, where our current design incurs unnecessary switches for cache-resident accesses, presenting future optimization opportunities. While Coro-AMU-D demonstrates performance comparable to software prefetching (even underperforming it at 100ns latency), the performance of CoroAMU-full with architectural enhancements validates the superiority of decoupling hardware.

\begin{figure}[hbpt]
    \centering
    \vspace{-1em}
    \includegraphics[width=0.48\textwidth]{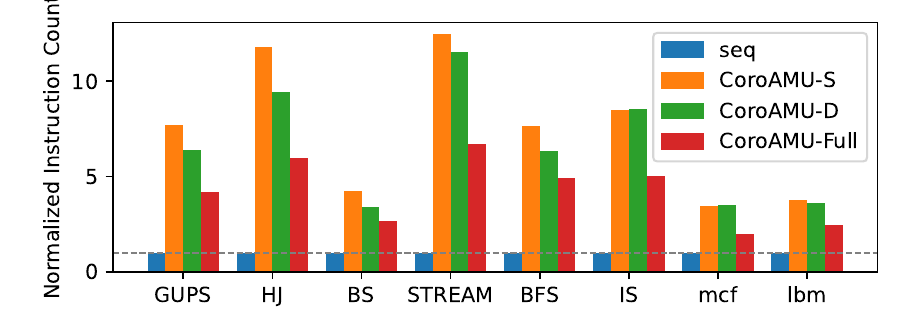}
    \vspace{-0.5em}
    \caption{Dynamic instruction count normalized to serial code, indicating extra control cost. The latency is 100ns.}
    \label{fig:eval_inst}
\end{figure}

\begin{figure}[hbpt]
    \centering
    \vspace{-1em}
    \includegraphics[width=0.48\textwidth]{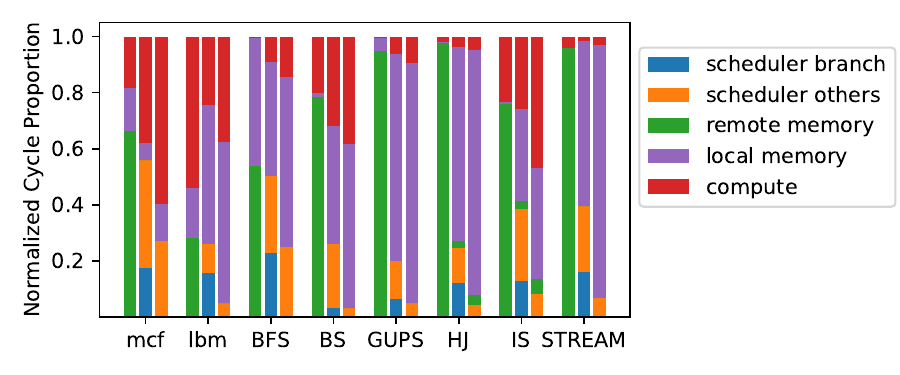}
    \vspace{-0.5em}
    \caption{Breaking down execution cycles for: (1) serial code, (2) CoroAMU-D, (3) CoroAMU-D with \texttt{bafin}. The latency is 200ns.}
    \vspace{-1em}
    \label{fig:eval_breakdown}
\end{figure}

\subsection{Performance Breakdown}

Coroutines embody a trade-off between control overhead and memory-induced stalls. While inserting control statements significantly increases dynamic instruction count, it concurrently elevates IPC, ultimately yielding performance gains. Figure \ref{fig:eval_inst} quantifies this instruction-to-control ratio: CoroAMU-S exhibits 6.70$\times$ dynamic instruction expansion, which CoroAMU-D reduces to 5.98$\times$ through hardware-managed SPM in L2 cache that eliminates software queue management. CoroAMU-Full achieves a 3.91$\times$ overhead via architectural extensions that encode coroutine metadata within memory operations and \texttt{bafin} instructions, completely offloading control flow to AMU hardware.
It serves as the primary contributor to performance gains while validating the necessity of software-hardware codesign.

Figure \ref{fig:eval_breakdown} further illustrates the control overhead through cycle breakdown analysis. The serial implementation spends substantial time on remote memory accesses. CoroAMU-D eliminates this latency by replacing all remote accesses with non-blocking decoupled instructions, instead introducing overhead from the scheduler and local memory operations (primarily context switching). Notably, branch mispredictions within the scheduler account for over 15\% overhead in CoroAMU-D on average, stemming from indirect jumps during coroutine switching. CoroAMU-Full eliminates it through \texttt{bafin} instructions. Yet, under 100ns latency, \texttt{lbm} and \texttt{STREAM} demonstrate better performance in their serial versions due to inherent memory locality, resulting in proportionally larger compute phase durations.

\subsection{Compiler Efficiency}

\begin{figure}[hbpt]
    \centering
    \vspace{-1em}
    \includegraphics[width=0.48\textwidth]{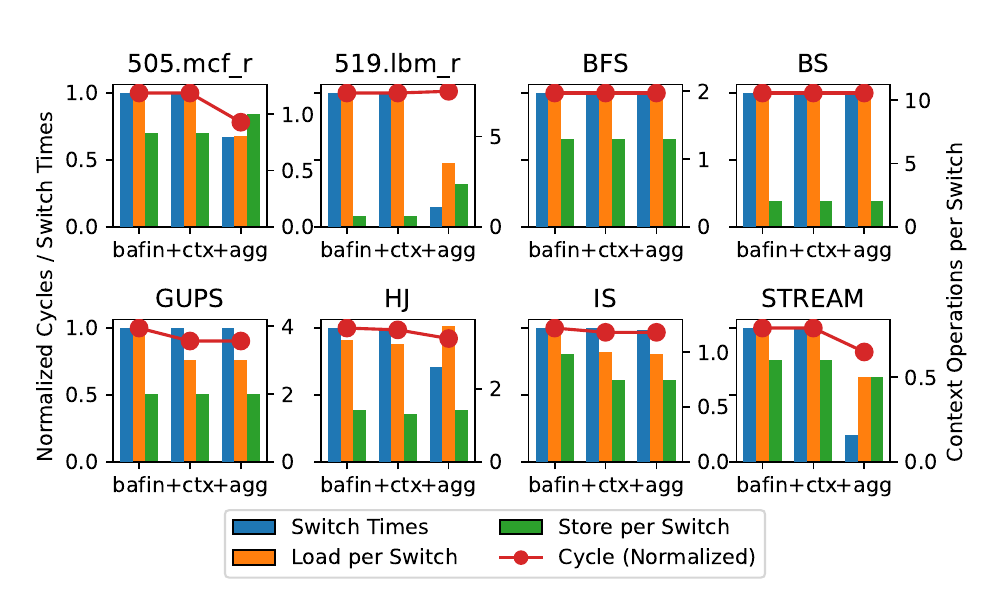}
    \vspace{-0.5em}
    \caption{Effects of compiler optimizations on performance (normalized), switch times (normalized) and context operations per switch: (1) CoroAMU-D with \texttt{bafin}, (2) Analyze variables to reduce context, (3) Request aggregation. The latency is 100ns.}
    \label{fig:eval_compiler}
\end{figure}

Figure \ref{fig:eval_compiler} demonstrates optimizations for reducing context switching overhead through two key mechanisms: context selection and request aggregation. The former optimization minimizes preserved registers per context, reducing load/store operations per switch, identified in \texttt{GUPS}, \texttt{IS}, and \texttt{HJ}. The latter decreases switching frequency while potentially increasing memory operations per switch, observed in \texttt{mcf}, \texttt{HJ}, \texttt{lbm}, and \texttt{STREAM}. The reduced switching overhead directly translates to performance gains, over 20\% at most. For \texttt{lbm}, the gain becomes evident under high latency, invisible at 100ns.

\subsection{Memory-Level Parallelism}

\begin{figure}[hbpt]
    \centering
    \vspace{-1em}
    \includegraphics[width=0.48\textwidth]{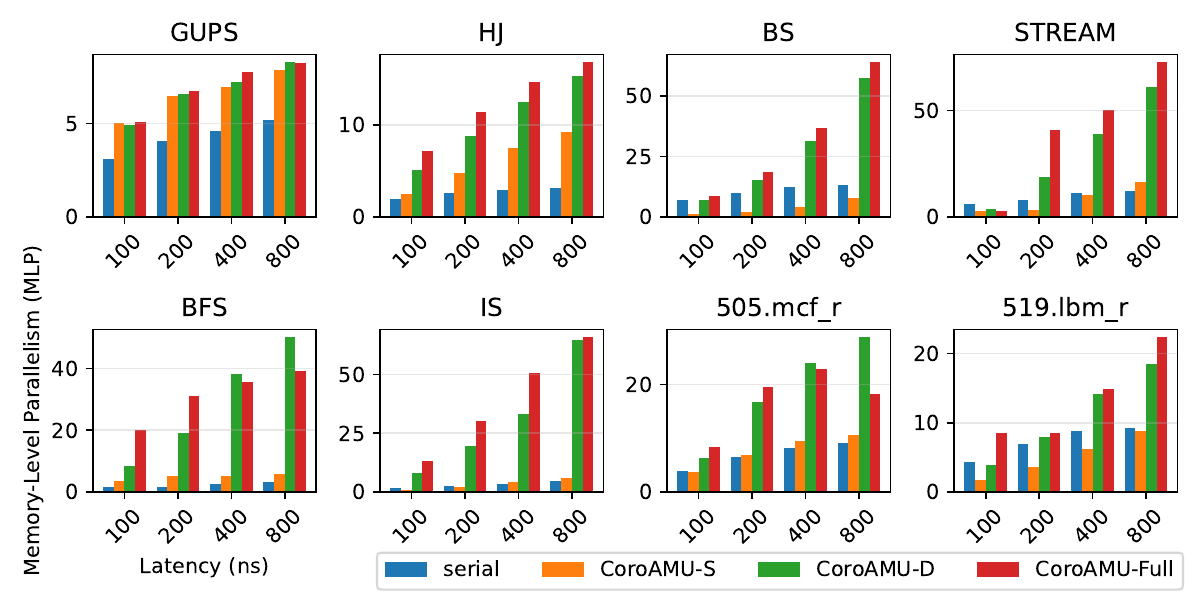}
    \caption{Memory-level parallelism for serial and CoroAMU, in terms of inflight memory requests at the memory controller.}
    \label{fig:eval_mlp}
\end{figure}

From the perspective of memory systems, latency-bound applications suffer from CPU stalls caused by memory accesses, resulting in underutilized memory bandwidth. This is one key optimization target of CoroAMU. Figure \ref{fig:eval_mlp} illustrates the average MLP during peak memory phases. For latency-sensitive applications like \texttt{GUPS} and \texttt{BFS}, the baseline achieves MLP of less than 5. While SOTA prefetching techniques reduce stalls, they remain constrained by core resources like MSHR entries with MLP capped below 20. Our decoupled memory access mechanism demonstrates superior scalability, achieving an MLP of 64 and revealing that existing MLP limitations become the performance bottleneck under high-latency conditions. 
This scalability could be extended by increasing the number of coroutines.

\section{Related Work}

\subsection{Decoupling and Interleaving}


Decoupled access-execute\cite{dae} was initially implemented as a CPU architecture approach, where memory access instructions and computational instructions are managed in separate queues. It has been replaced by modern out-of-order processors with large reorder buffers (ROBs), but the underlying concept is now increasingly being adopted by compilers for software optimization. DAE compilers\cite{jimborean2018cgo}\cite{decouple_efficient} cluster load instructions and split programs into two course-grained phases. It is improved by multi-versioned, fine-grained, and prefetching phases\cite{daedal}, and across-phase instruction scheduling\cite{clairvoyance}. Minor hardware support of cache check-miss further enhances this approach for in-order cores\cite{decouple_swoop}. 

Programming models such as finite state machines\cite{amac} or coroutines\cite{prefetch_sw_coro}\cite{decouple_cimple} can manage the progress of every load instruction in access phases individually, which is coverd in Sec. \ref{sec:background}.

Helper threads are another way of multi-threading to improve single-core performance. They leverage thread-level parallelism in a single core to utilize idle computational resources when one thread is blocked, thereby enhancing throughput. These threads could perform speculative precomputation\cite{runahead_specprecomp}\cite{helper_spec_predict}, or prefetching\cite{helper_inter_prefetch}\cite{helper_prefetch_precomp}, with the support of either software schedulers\cite{helper_soft}\cite{helper_vmt} or hardware such as SMT\cite{smt} and customized messaging queues\cite{helper_thread}. Decoupled Vector Runahead\cite{decoupled_vector_runahead} employs a fully autonomous hardware thread to enable aggressive prefetching for the main thread. Compared with other multi-threading models, helper threads provide few benefits for highly memory-intensive programs\cite{helper_limit} and lack scalability and stable performance advantages for large-scale applications facing increasing latency.

\subsection{Branch Misprediction Elimination}

For simple if-else control flows and value selection, predication\cite{branch-predicate}\cite{branch-predicate-opt} is an effective means of avoiding branch prediction, converting a branch into a computation instruction. Early resolution\cite{branch-early-resolution} analyzes control flow and computes branch outcomes in advance, providing accurate information for branch prediction. Control-Flow Decoupling\cite{multi-ctrlflow}\cite{branch-decouple} further uses a queue to transfer branch results. Software approaches heavily depend on compiler transformation, limiting their applicability.


Probabilistic branch support\cite{branch-prob} is a mechanism for eliminating mispredictions for random numbers, similar to the main idea of \texttt{bafin}, adding a 4-entry prediction table to the BPU, and using the previous iteration's random number results to guide current branch prediction. It only applies to random algorithms and maintains statistical properties rather than the correctness of every execution.

\subsection{Parallelization and Multi-core Architecture}

In high-performance computers, leveraging multi-core processors has become a fundamental strategy. Existing approaches for accelerating irregular applications typically employ advanced parallelization techniques such as software pipelining \cite{multi-dsp}\cite{multi-dsp-hard}\cite{multi-dsp-fine} and control-flow decoupling \cite{multi-ctrlflow} to expose program parallelism and effectively distribute workloads across multiple cores. Modern compiler frameworks frequently utilize polyhedral model-based analysis \cite{multi-poly2010}\cite{multi-poly2012}\cite{multi-static} to systematically identify and extract parallelized loop components. Architectural innovations like HELIX \cite{multi-helix}\cite{multi-helixrc} demonstrate the potential of architecture-compiler co-design for automated parallelization and minimizing multi-core communication costs. Thread-level speculation techniques \cite{multi-posh}\cite{multi-t4} provide additional opportunities to enhance scalability for complex workloads.

CoroAMU will complement parallelization approaches in the future, maintaining compatibility with multi-core programming models while achieving more efficiency by combining the benefits of both inter-core and intra-core parallelism.

\section{Conclusion}

We highlight critical limitations
in existing approaches for coroutine-based memory latency hiding. 
While coroutines effectively mitigate memory access stalls, our analysis reveals that their runtime overheads remain nontrivial. Furthermore, traditional static schedulers relying on software prefetching increasingly struggle to adapt to emerging disaggregated architectures. 
To address these challenges, we present CoroAMU, featuring a compiler for memory-centric coroutines, as well as an easy-to-use OpenMP-style programming interface. By enhancing the AMU hardware with coroutine-tailored improvements to its decoupled issue-poll memory interface, we demonstrate the superiority of the dynamic scheduling mechanisms of coroutines. Experiment results show that CoroAMU achieves a remarkable 3.39$\times$ speedup over baseline processor at 200ns latency. These results underscore the importance of hardware-software co-design in unlocking the full potential of coroutines for modern memory systems.

\section{Acknowledgements}

We sincerely thank the anonymous reviewers for their insightful suggestions. This work is supported by the National Key R\&D Program of China (Grant No. 2022YFB4500403).


\bibliographystyle{IEEEtran}
\bibliography{sample-base.bib}

\end{document}